**Influence of chain topology and bond potential on the glass transition of polymer chains simulated with the bond fluctuation model**

Juan J. Freire

Departamento de Ciencias y Técnicas Fisicoquímicas, Facultad de Ciencias, Universidad Nacional de Educación a Distancia (UNED), Senda del Rey 9, 28040 Madrid, Spain

ABSTRACT

The bond fluctuation model with a bond potential has been applied to investigation of the glass transition of linear chains and chains with a regular disposition of small branches. Cooling and subsequent heating curves are obtained for the chain energies and also for the mean acceptance probability of a bead jump. In order to mimic different trends to vitrification, a factor $B$ gauging the strength of the bond potential with respect to the long-range potential (i.e. the intramolecular or intermolecular potential between indirectly bonded beads) has been introduced. (A higher value of $B$ leads to a preference for the highest bond lengths and a higher total energy, implying a greater tendency to vitrify.) Different cases have been considered for linear chains: no long-range potential, no bond potential and several choices for $B$. Furthermore, we have considered two distinct values of $B$ for alternate bonds in linear chains. In the case of the branched chains, molecules with different values of $B$ for bonds in the main chain and in the branches have also been investigated. The possible presence of crystallization has been characterized by calculating the collective light scattering function of the different samples after annealing at a convenient temperature below the onset of crystallization. It is concluded that crystallization is inherited more efficiently in the systems with branched chains and also for higher values of $B$. The branched molecules with the highest $B$ values in the main chain bonds exhibit two distinct transitions in the heating curves which may be associated with two glass transitions. This behavior has been detected experimentally for chains with relatively long flexible branches.

**Introduction**

Vitrification is a process associated with fast cooling of liquids and mainly characterized by a loss of mobility which is not caused by a long-range organization of the molecules. In these conditions, the system becomes an amorphous solid or glass.[1] Vitrification does not correspond to a thermodynamic transition and, in fact, a glass is not in thermodynamic equilibrium. However, the process usually takes place in a narrow range of temperatures that permits the definition of an experimental glass transition temperature, $T_g$. The glass formation can be observed at slower cooling rates in some types of materials as polymer melts,[2] whose crystallization is partially or totally hindered.

The glass transition of polymer melts has been the object of several studies by numerical simulation in recent years.[3-11] Given the large time scales involved in the process and the need for considering a relatively large portion of the molecular systems, a substantial part of these studies have made use of simplified coarse-grained polymer models. Some of these studies have used the bond fluctuation model,[12-13] devised to offer the computational benefit of lattices and the more realistic characterization of systems represented by the open space simulations. This model offers a variety of bond lengths. Also, it permits the generation of successive configurations in a Monte Carlo (MC) simulation by means of a simple single bead jump. These features make this model adequate for the study the statics and dynamics of vitrification when intramolecular and intermolecular interactions are included by means of adequate temperature-dependent potentials. A long-range potential that describes interactions between non-bonded close units is usually introduced. However, this potential may eventually induce the system crystallization if the mobility of the polymer units allows for the formation of long-range order. Vitrification takes place, however, at a lower temperature when mobility of the amorphous structure is severely limited.

Previous studies with the bond fluctuation model have demonstrated that vitrification can only be investigated with the additional introduction of a bond-length potential.[3,6,10] This potential leads to a higher proportion of the longest bonds allowed in the model, which are able to trap some amount of empty space, reducing the system mobility and preventing the formation of ordered structures. In some of these studies, other intramolecular or intermolecular interactions are completely removed to avoid crystallization[3] so that the temperature variation is only reflected in the bond potential effectiveness. A recent work on this issue has concluded that the bond potential should actually be much stronger than any long-range potential in order to mimic a realistic glass behavior.[10] This particular conclusion was achieved by using a fixed bond potential and several weakened forms of the long-range potential. Previous studies of the glass transition with the bond fluctuation model have been focused in the study of linear chains.

This work considers linear chains and also chains with a regular disposition of small branches (which we denote as "branched chains") represented by the bond fluctuation model in order to make a more systematic study of the possible of crystallization in glass forming polymer melts. Most of the systems have a fixed long-range potential, which consistently leads to the same range of temperatures for the possible crystallization of the system, and differ in the strength of the bond potential and, also, in the polymer topology. Furthermore, we include in our study the cases of chains with different assignments of bond potentials within the chains (alternate bond potentials for linear chains, and different potentials in the main chain and branches in the case of non-linear chains). Crystallization is mainly characterized by the increase of the collective scattering function of the simulation box.

**Model and Methods**

The bond fluctuation model [12,13] offers a considerably higher number of empty positions available around an occupied site and, furthermore, it only needs a single and simple form of elementary move, particularly useful when it comes to writing codes applicable to non-linear chains. It considers mutually- and self-avoiding walks on a cubic lattice, each polymer bead effectively occupying eight corners of a unit cell.

In the present work, the distance between adjacent lattice sites is adopted as the length unit and energies are given in $k_BT$ units. We use a version of the model with an attractive-repulsive potential for bond energies, $V(l)$, depending on each particular bond length, and a long-range potential $U(R)$, depending on the particular distance between non-neighboring units. Namely, we consider the potentials proposed by Wittkop et al.[6] to mimic the "glassy" behavior of polymers chains. Our version, however, introduces a numerical factor for the bond energy potential energy, $B$, with respect to the values employed by Wittkop et al. so that

$$V_B(l) = BV(l) \qquad (1)$$

The consideration of factor $B$ is inspired in recent results suggesting that the main features of the glass state can be only reproduced if the strength of the bond potential is somehow increased with respect to the long-range potential.[10] The present study, however, differs from the investigation performed in Ref. 10 in some substantial features. Firstly, a factor $B$ is introduced here to increase the absolute values of the bond potential. Secondly, the temperatures $T$ actually correspond to the factors, $k_BT$, that determine the Boltzmann weights associated to the potential energy terms, whose absolute values are presented in Tables 1 and 2 of Ref. 7 and in Table 1 of Ref. 14. These two differences apparently mean that the temperature values in this work correspond to temperatures $(1+B)^2$ times smaller than in Ref. 10, although this correspondence is not exact due to some differences in the specific forms of the potentials. We have considered different values of $B$. In particular, we have investigated the case $B=0$ (no bond potential) and the range $B=1-5$. We have also obtained some results without the long-range potential and $B=1$.

Furthermore, our particular model considers linear chains and branched chains that are built by adding a single unit branch to every non-end bead in a linear chain. The main-chain branching beads are treated as "chiral atoms". Consequently, we consider the plane formed the two bonds in the main chain connected to every branching bead (when these two bonds are not in a collinear configuration) and define the side of this plane in which the first bead of the branch should remain along the simulation. The initial spatial disposition is assigned for each unit branch during the initial equilibration process. We have verified that substantial amounts of "meso" and "racemic" diads are randomly generated by using this procedure.

The model includes the possibility of employing different values of $B$ for different types of bonds. In particular, the case of different $B$ factors on alternate bonds along a linear chain has been considered. This case can mimic some chains with a regular disposition of monomers (or larger groups of atoms) with different compositions and it may also apply to the description of some polymers with different side groups along the main chain structure. Also, we have performed simulations for our branched chains considering the cases of a different $B$ factor (greater or smaller) for bonds in the branches with respect to the factor associated to bonds in the main chain.

The systems are built by placing $n$ chains, each composed of $N$ beads, in a cubic lattice of fixed length $L$, with periodic boundary conditions.[15] Therefore the simulations correspond to a NVT ensemble. The value of $L$ is chosen to be high enough so that the number of interactions between different replicas of the same chain is very small, $L \geq 2<R^2>^{1/2}+5l$, where $<R^2>$ is the mean quadratic end-to-end vector of a single self-avoiding walk chain and $l$ is the root-mean squared bond distance in the absence of any type of numerical potentials ($l$=2.72). In the present work $L$=84 for all the cases. The number of chains is determined from the desired number of sites blocked by polymer beads, equivalent to the polymer volume fraction, $\Phi=8nN/L^3$. This volume fraction is fixed to the value $\Phi$=0.5, which gives a good representation of the melt state. Linear chains are constituted by $N$=40 beads, while branched chains are composed of $N$=78 beads, 40 of them form the main chain. Therefore, the non-end beads of the main chain define 38 branching points, which are connected with 38 single bead branches.

Initially, the chains form a packed configuration. Linear or branched chains are organised in successive arrays of extended conformations, placed as close as possible to avoid overlapping, i.e. at a distance of 2 for the linear chains. The separations between branched chains are 2 or 4 at the two different directions perpendicular to the main chain. The wider separation in a given direction allow for the location of the branch beads, bonded to the main chain in this particular direction. All initial bond lengths are 2. These initial configurations are far from equilibrium but the simulation boxes contain a large empty space in contact with some of the chains which eventually allow for a fast equilibration as it has been previously verified for linear[15] and star[16] chains.

An equilibration run of $10^6$ MC steps is performed to equilibrate the linear and branched chains without intermolecular or bond potentials, starting from their initial extended conformations. We have computed the mean quadratic radius of gyration in the last $2 \times 10^5$ steps, verifying that it is stable. Furthermore, we have compared its value, $<S^2>=78.5\pm0.5$, with the result obtained with a dilute (single chain) system at temperature, $T=4.67$, close to the θ state (same number of configurations), $<S^2>=82.5\pm0.5$. Both values are close, as is expected, since the chains are supposed to behave ideally in a melt and in the theta state. A slight difference between both values is also found for linear chains, $<S^2>=61.6$ in the melt and $<S^2>=66.3$ at $T=4.67$, as was reported in a previous study of conformational and dynamic properties with the same model.[15] These data confirm that this equilibration method (starting form extended conformations as described above) can be employed correctly also for the apparently more difficult case of branched chains.

A second equilibration run, also of $1 \times 10^6$ steps is performed at a conveniently high temperature ($T=20$) for every system with a different set of potentials. Starting at this point, we perform cooling runs where the system temperature is gradually decreased. We have considered several cooling rates, corresponding to $\Delta T=-0.1$ per 5, 50, 500, $10^3$, $5 \times 10^3$, $10^4$ and $2.5 \times 10^4$ MC steps. The runs are finished when a very low temperature ($T=0.1$ or $0.2$) is reached. "Isothermal annealing" processes are subsequently simulated at different low temperatures, using the configurations obtained from the cooling process. Most of these runs are of similar length to the equilibration processes, though sometimes they are significantly extended (up to $3 \times 10^6$ MC steps) in subsequent runs in order to try to detect the possible formation of intermolecular structures. Furthermore, heating runs are also performed from the lowest temperature of the cooling processes, with the same rates employed for cooling.

In order to detect intermolecular order, we have computed the collective scattering function of the systems. This function is obtained as

$$S_{coll}(q) = 8 n_s^{-1} < \sum_{i}^{n_s} \sum_{j}^{n_s} f_i f_j \, exp(i\mathbf{q}.\mathbf{R}_{ij}) > \qquad (1)$$

In eq 1, $q$ is the scattering vector, depending on experimental settings, the vectors $\mathbf{R}_k(t)$ refer to the positions of the different $n_S = L^3$ sites within the system and $f_k$ is the contrast factor, related with the difference between the scattering factor due to the particular occupation in the site in a given configuration and the mean scattering factor of the system. Thus, for the homopolymer systems

$$f_i = 1 - \Phi/8 \qquad (2a)$$

if site $i$ contains a bead unit.

or

$$f_i = -\Phi/8 \qquad (2b)$$

otherwise, in order to comply with the requirement that the global system, considered as a large single isotropic volume, does not scatter.[15] The practical range of values of $q$ is limited by the periodic boundary condition in the simulation box:

$$q_k = (2\pi/L)n_k, \qquad k \equiv x,y,z, \qquad n_k = 1,2..., \qquad (3)$$

which also determines the smallest value of *q*. With these specifications, a homogeneous system should show small and isotropic values of $S_{\text{coll}}(q)$ for the whole range of small and intermediate values of *q*. (At large values of *q* local features of the model can be observed). In the presence of crystallization, however, $S_{\text{coll}}(q)$ becomes large and anisotropic for small values of *q* (long-range order).

**Results and discussion**

In Figure 1 we show the mean energy per bead, *E*, along cooling processes corresponding to six different cases. These curves have been obtained at a cooling rate of $\Delta T$=-0.1 per $5 \times 10^3$ MC steps. We consider linear and branched chains with bond potential, *B*=1, or without this contribution (*B*=0). We also include the results corresponding to systems without long-range potential (linear or branched chains) but with bond potential (*B*=1). All these curves exhibit a clear sigmoidal form, as has also been observed in previous simulations with the same model.[6] Considering only the systems with long-range potential, the curves with *B*=1 have lower energies than the curves with *B*=0 since the bond potential gives a negative contribution to the energy. The curve corresponding to the linear chain without bond potential clearly shows the most abrupt transition. Comparing the curves with the same type of potential, the sharpest transitions always correspond to the linear chains. (This effect is, however, slight for the systems without long-range potential). Although the form of the curves may indicate a first order transition, as crystallization, the downwards curvature at high temperatures is a feature that can simply be associated to the asymptotic behavior of the energy at high temperatures. This behavior is implicit in the model (the sigmoidal shape is also observed in the absence of long-range interactions, case for which crystallization does not occur.[3,4]) At very high temperatures, all the different chain configurations have similar statistical weights. Therefore, a further increase of temperature does not have any effect on the distribution of configurations and, consequently, the mean energy stays constant. The asymptotic regime is reached at higher temperatures when the total energy is greater and, consequently, the presence of a more energetic bond potential implies a later arrival of the asymptotic behavior, though this behavior should eventually be reached at conveniently high temperatures. Only simulations in the NPT ensemble[13,17] would be able to exhibit an increase in the number of vacancies in the system at higher temperatures implying an increase of energy (decrease of favorable interactions) in this region. Although some algorithms have been designed to perform non-constant volume simulations in lattices, these algorithms have not been implemented yet for the bond fluctuation model.

However, the more abrupt curves exhibited by some of the curves obtained with long-range potential suggest a substantial crystallization process. This seems to explain the differences between the curves corresponding to the linear and branched chains that share the same asymptotic limits at high temperature. One may also suspect that crystallization explains the more marked sharpness in the curves of the models without a bond potential. In fact, the bond potential has been included in the bond fluctuation as an artifact to avoid crystallization. It can be observed that the centers of the sigmoidal transitions for all the curves corresponding to the systems with long-range potential in Figure 1 are placed at similar temperatures. In fact, these curves are sharp enough to give a good estimate of a transition temperature associated to the center of the crystallization process, $T_c$. Their first derivatives provides $(T)_c=3.3\pm0.1$. The systems without long-range interactions, however, show an inflexion point at considerable lower temperature which indicates a different physical behavior.

Actually, the onset of the crystallization process can actually be assumed to occur at considerable higher values of $T$. However, it is more difficult to characterize since it is somehow masked by the asymptotic bending of the curves. This feature can also be discussed by taking into consideration other studies for the bond fluctuation model with the same type of long-range and bond potentials for polymer-solvent systems. As previously mentioned, the theta temperature of this particular model (linear chains) was estimated to be close to $T=5$.[14,15,18] According to the standard theory for polymer solutions,[19] this temperature corresponds to the critical point of infinitely long polymer chains. For finite chains, the critical point should occur at slightly lower values of $T$. All the present models show critical temperature located at the maximum in the phase-separation temperature-concentration curve. At phase separation, a polymer-solvent system yields a polymer-rich solution and a more dilute phase. Since solvent molecules are actually represented by the model vacancies, polymer-rich solutions are equivalent to the systems of pure polymer with vacancies investigated in the present work. Therefore, the high concentration region of the phase separation curve can also mark the initial steps of the formation of a large crystalline region at a temperature slightly below $T_\theta$.

Since all curves shown in Figure 1 are sigmoidal, the characterization of the glass transition in the low temperature region is not easy. This transition is generally marked by an abrupt change in the slope of the curves, associated with a discontinuity (not a peak) in the first derivative (specific heat). The curves in Figure 1 suggest two regions at low temperature. In one of these regions, the value of the energy is practically constant (first derivative close to zero). The second region corresponds to a relatively narrow range of temperatures where the curve shows a constant slope. This region is broader for the systems for which we can expect more inhibition of crystallization (those including the bond potential or branched chains). The intersection of the slopes corresponding to these two regions may be used as an estimate of the apparent glass transition temperatures (see Table 1). It is observed that the branched chains show smaller values of $T_c$, in spite of their higher number of beads and the steric effects of the branches. This has also been verified for the systems without long-range potential, where, however, the differences between the systems with linear and branched chains are small.

We have investigated whether similar conclusions can be achieved by making a representation of the mean acceptance probability for a bead jump, $p$, obtained with the cooling processes, shown in Figure 2. This average measures the possibility to perform the jump and, therefore, is related with the number of empty sites available around a bead and also on the Metropolis probability to perform a bead jump to these sites. These properties have been proposed or used to characterize local changes in the specific volume that are also indicative of a glass transition.[6,10,11,20,21] We can actually observe some correlation in the features shown by the equivalent curves in Figures 1 and 2, though the jump acceptance curves are markedly less sharp. In any case, our estimation of the glass transition temperature will be based on the energy curves, since these graphs provide a clearer distinction between the different cases, especially at low temperatures, where the acceptance ratios become very small.

Rigidity is an important factor determining the glass transition temperature of polymer systems.[2] In fact, in the comparison between real linear chains and chains with the same skeleton but with bulky side groups, the latter have a higher $T_g$ because of the considerably higher rigidity of the main chain induced by the moiety. In order investigate the rigidity due to branching in the present model we have calculated the mean quadratic distance between the end units of the backbone for the melt of branched chains without potential (infinite temperature). The result is $<R^2>=460\pm2$ for the last $2\times10^5$ steps of the initial equilibration. This result can be compared with the slightly smaller value for the mean quadratic end-to-end distance of linear chains (same number of beads in the backbone) previously reported, $<R^2>=370$. The conclusion is that the increase of rigidity due to branching is relatively small (a 12% increase in the characteristic ratio).

In Figure 2 it can also be observed that the acceptance ratios are higher for the linear chain than for the equivalent (same model) branched chains at high temperatures. Consequently, for a similar main-chain length, the higher rigidity of the main chain in the branched chains seems to be the predominant factor to inherit bead jumps. However, the branched chains have considerable higher acceptance ratios in the low temperature range. In this case, the much higher number of chain ends with a higher mobility in the branched chains turns to be the mean factor. This effect is found in spite that the translational diffusion coefficients, $D$, (obtained from representations of the mean squared global displacement of individual chains vs. number of MC steps) are always considerably greater for the case of linear chains. For instance, comparing the systems with long-range potential and $B=1$ at the intermediate temperature $T=5$, the acceptance probability is larger for the branched chains. Nevertheless, $D=5\times10^{-6}$ for the linear chains and $D=2\times10^{-6}$ for the branched chains at this temperature. Since the glass transition occurs in the low range of temperatures, the mobility of end beads explains the lower $T_g$ values exhibited by the branched chains. Actually, it has been experimentally verified for (similarly rigid) polyethylene chains of different degrees of branching that $T_g$ decreases with a variable summing up the number of branching units and monomers in the branches.[22] Incidentally, the acceptance ratios of the linear chains are again higher at very low temperatures (glass state) in the case of the chains without long-range interactions. The early occurrence of the glass transition for this case seems to trap a larger amount of empty space allowing more local jumps than in the case of branched chains.

In Figure 3 we show the cooling curves of a branched chain with $B=1$ obtained with different cooling rates ($\Delta T=-0.1$ per 5, 50, 500 and $5\times10^3$ MC steps). All these curves show similar sigmoidal trends. Faster cooling processes are obviously associated to less abrupt overall transitions. The first derivatives show broader peaks for the curves corresponding to the faster cooling processes which complicate the characterization of a transition temperature. However, all these peaks are compatible with our estimate of crystallization temperature, $T_c \cong 3.3$. Assuming that the glass transition can be characterized by the point where the slopes corresponding to the two linear regions at low temperature intercept, we can observe that a fastest rate curves correlate with the lowest glass transitions (see Table 1). This tendency cannot be justified except by considering that the $T_g$ estimates are strongly biased by the simultaneous crystallization processes. The slowest cooling processes contain larger crystalline regions acting as effective crosslinks. Therefore, the amorphous parts of the system have less mobility, increasing the glass transition temperature.

In Figure 4, we show the results obtained for the branched chain with different values of $B$ for the cooling rate of $T$=-0.1 per $10^3$ MC steps. In order to compare more easily the different curves, the energy values are divided by the factor $(1+B)$. It can be clearly observed that higher values of $B$ tend to retard the asymptotic regime. As a consequence, the glass transition temperature is increased and becomes closer to the crystallization temperature range. Therefore, crystallization can effectively be inhibited or even eliminated by increasing $B$. From this point of view, the value $B$=5 could be adequate to give a particularly good description of the glass transition.

The estimation of the glass transition from these three curves is also given in Table 1. The result for $B$=5 is considerably higher than the center of hypothetical crystallization temperature range, assuming that the latter is mainly determined by the long-range potential and, therefore, it should be close to our estimation of $T_c \cong 3.3$ for $B$=0 or 1. A useful model should be able to provide ratios between crystallization and the glass temperatures following the main trend of most real polymers, for which vitrification occurs at temperatures significantly lower than possible crystallization. Otherwise, crystallization is not possible from the kinetic point of view, as it actually happens in the case of our model with $B$=5. Therefore, our choice $B$=3 can be descriptive of these realistic polymer systems. Even though the curve for $B$=3 shows a sigmoidal shape at high temperatures, the slope is constant in the range of temperatures for which crystallization may occur.

It should also be mentioned that the acceptance probabilities of the branched chains are above the values corresponding to the equivalent linear chains in the whole range of temperatures above glass transition for the models with $B>1$. Therefore, higher bond energies seem to stress the effect of the higher mobility of end units with respect to rigidity in the determination of the average bead jump probability for branched chains.

In Figure 5 we show the results for a branched chain with $B=3$ obtained at different cooling rates. ($\Delta T = -0.1$ per 5, 50, $10^3$, $10^4$ and $2.5 \times 10^4$ MC steps). It can be observed that the estimate of $T_g$ from these curves is practically constant. It should be remarked that the two curves obtained with the slowest cooling rates exhibit remarkably close and constant slopes in the region above $T_g$, showing a near to equilibrium curve in this region. The results for $B=5$ at different cooling rates yield similar conclusions. Therefore, the marked influence of crystallization in the estimation of $T_g$ observed for the $B=1$ results is eliminated when $B$ is increased, which can be considered as an indirect verification of the lack of significant crystallization in the $B>1$ systems.

A more direct proof of the inhibition of the crystallization process for higher values of $B$ can be done by computing the scattering function, $S_{coll}(q)$. This function has been previously used to investigate the local structure (large $q$ values) and the isothermal compressibility (low $q$ range) in the melt and glass state for polymer systems without long-range potential.[4,13] The collective scattering function for athermal polymer systems exhibits a flat form close to the value $S_{coll}(q) \cong 0.2$ at low $q$.[13,23] In our case, we are interested to characterize the possible presence of crystallization in the investigated systems. The onset of this feature should induce the appearance of peaks in the region of lower values of *q* due to the presence of long-distance order. These peaks should eventually lead to a divergence at q→0 marking the spinodal of the transition process. In Figure 6, we present the curves corresponding to results obtained in different cases obtained at the lowest temperature of the cooling process (*T*=0.2). Ordered structure is observed for the systems with *B* smaller than 3. The most prominent peaks at low *q* correspond to the chains without bond potential. For *B* =1, the peaks are significantly smaller, but they are still clearly observed. However, the curves are mainly flat for *B* =3. For a given value of *B*, the relevant peaks are noticeably smaller in the case of branched chains. This shows a considerable inhibition of crystallization in our branched chains.

The features marking long-range order should progressively be enhanced in an annealing process at a temperature lower than the onset of the crystallization process but higher than the glass transition temperature, where the crystals are able to grow at an adequate rate. In Figure 7, we show the results obtained for the model with *B*=1 (linear and branched chain) with an annealing at *T*=3.7, value close to the estimated temperature for the center of the crystallization transition. We observe a great increase of the intensity of the peaks, correlated with their progressive displacement to lower values of *q* for the linear chains. The increase is less dramatic for the branched chains, confirming the important inhibition of crystallization in these systems. However, ordered structures are eventually formed even for this type of chains, unless *B* is greater than 1. For higher values of *B*, we have mainly investigated this feature in linear chains, since their tendency to crystallize is higher. In Figure 8, we observe that these features are significantly less marked for *B* =3. In fact, the system seems to exhibit some long-range inhomogeneity at the earlier steps of the annealing, but the rate of this effect is retarded and is practically stopped at longer times. An even clearer inhibition in the ordering growth is found for a linear chain with the alternative choices of $B=1$ or $B=5$ for odd and even bonds along the chain skeleton, showing remarkably small values of the scattering functions after many annealing steps, see Figure 9. This choice of values of $B$ tries to mimic the presence of bulky (prone to vitrify) monomers in the polymer structures.

In Figure 10, we present the cooling curves of $\Delta T=-0.1$ per $10^3$ MC steps corresponding to different types of chains, all of them with a mean value of $B$ close to 3. Comparing the curves for a linear and a branched chain, both of them for $B=3$, it is observed that the linear chain exhibits a slightly sharper overall transition. The estimated value of $T_g$ for the linear chain is higher ($T_g \cong 3.5$) though it is still below the onset of a possible crystallization transition. We include in Figure 10 the cooling curve for the linear chain mixing $B=1$ and $B=5$ bonds. This curve exhibits a considerable large region with constant slope above $T_g$ so that sigmoidal shape is only observed when high temperatures (T>10) are considered. This is consistent with the strong inhibition of crystallization shown by the scattering curves. Its glass transition is estimated to be slightly smaller than the result for linear chains with a single $B=3$ value (Table 1).

Also, we include cooling curves that correspond to branched chains where either the main chain or the branch bonds have more tendency to form a glass, $B=5$, while we assign $B=1$ to the other chain bonds. The curve corresponding to "glassy" units in the branches is similar to that corresponding to linear chains with alternate $B$ values. However, the chains with a "glassy" main chain are substantially different, with a broader linear region above $T_g$, as it can be observed in Figure 10. This region extends up to T≅12, though a slight change in the slope can be distinguished in this region, at $T_t$≅6. Moreover, the most significant change in the slope yields an apparent glass transition at very small temperatures, $T_g$≅1.2, close to the smaller $T_g$ values obtained for $B$=1 with higher cooling rates (which supposedly may inhibit the effect of crystallization on the determination of $T_g$ for this value of $B$) or in the absence of long-range interactions.

We have obtained heating curves with the same rate for the variation of $T$ used for cooling. Heating starts at the lowest $T$ reached in the cooling process without annealing. The results for the branched chains with more glassy branches are shown in Figure 11. There is some hysteresis above $T_g$, and both curves converge to a common equilibrium asymptotic behavior at large values of $T$. A similar description applies to the rest of investigated systems except for the branched polymers with a more glassy main chain. The peculiar behavior of these polymers is verified in the analysis of the heating curve. The cooling and heating curves for this particular system are shown in Figure 12. The heating curve above shows two clearly different regions above $T_g$ that correlate with the slight differences in slope along the cooling curve. The first region, between $T_g$ and $T_t$ is similar to the curves observed in other systems, with a slight hysteresis. However, at temperatures above $T_t$, the heating curve bends downwards and the hysteresis is significantly increased. Since our estimates for $T_g$ and $T_t$ roughly correspond to the glass transition temperatures of chains with $B$ =1 and $B$ =5, respectively, the conclusion is that these systems show local heterogeneity, with two regions of independent glass transitions.

Real polymers that may correspond to this model are chains with bulky main chain monomers and flexible branches. In particular, these simulation results seem to be consistent with dielectric relaxation[24] and NMR[25] experiments performed for poly(di-n-alkylitaconates) which apparently exhibit two glass transitions. It must be pointed out that the distinction between the cooling and heating processes due to hysteresis is only fainted noticed in the smother jump acceptance probability vs. $T$ curves for all the investigated cases. Due to this lack of resolution, the acceptance curves are not able to reveal the peculiar behavior of the system with a glassy main chain. However, these curves also again considerably smaller glass transition temperatures with respect to the system with glassy branch units.

In summary, we have confirmed the usefulness of employing stronger bond potentials to provide a better description of the glass transition in models that include a long-range potential. The systems with branched chains show a clear inhibition of crystallization and slightly smaller glass transition temperatures with respect to the systems of linear chains. The use of two different sets of bond potentials within a given type of chains is useful to describe more complex systems, as the case of branched chains with a more rigid main chain and flexible branches, for which two different glass transitions have been experimentally detected.

**Acknowledgement**

This work was supported by Project CTQ2006-06446 from DGI-MEC Spain. The author thanks Professor Arturo Horta (Universidad Nacional de Educación a Distancia, Spain) and Professor José Luis Gómez Ribelles (Universidad Politécnica de Valencia Spain) for useful discussion and also to Dr. Carl McBride, (Instituto de Química Física) Rocasolano, CSIC, Spain) for revising the manuscript.

**Table I.** Estimation of the glass transition temperature for different systems and cooling rates.

| System | $T_g$ |
|---|---|
| linear chains, $B=0$ | $2.4\pm0.2^a$ |
| branched chains, $B=0$ | $2.0\pm0.2^a$ |
| linear chains, $B=1$ | $2.2\pm0.2^a$ |
| branched chains, $B=1$ | $1.8\pm0.1^a$ |
| linear chains, $B=1$, $U(R)=0$ | $1.1\pm0.1^a$ |
| branched chains, $B=1$, $U(R)=0$ | $1.0\pm0.1^a$ |
| branched chains, $B=1$ | $1.7\pm0.1^b$ |
| branched chains, $B=1$ | $1.5\pm0.1^c$ |
| branched chains, $B=1$ | $1.1\pm0.1^d$ |
| linear chain, $B=3$ | $3.5\pm0.2^e$ |
| branched chains, $B=3$ | $3.0\pm0.2^e$ |
| branched chains, $B=5$ | $4.5\pm0.2^e$ |
| linear chains, alternate bonds[f] | $3.2\pm0.2^g$ |
| glassy branches[h] | $3.3\pm0.2^g$ |
| glassy main chain[i] | $1.2\pm0.1^g$ |

[a]cooling rate: $\Delta T=-0.1$ per $5\times10^3$ MC steps; [b]cooling rate: $\Delta T=-0.1$ per 500 MC steps; [c]cooling rate: $\Delta T=-0.1$ per 50 MC steps; [d]cooling rate: $\Delta T=-0.1$ per 5 MC steps; [e]all cooling rates; [f]$B=1$ or $B=5$ for odd and even bonds; [g]cooling rate: $\Delta T=-0.1$ per $10^3$ MC steps; [h]$B=5$ for bonds in the branches, $B=1$ for bonds in the main chain; [i]$B=5$ for bonds in the main chain, $B=1$ for bonds in the branches.

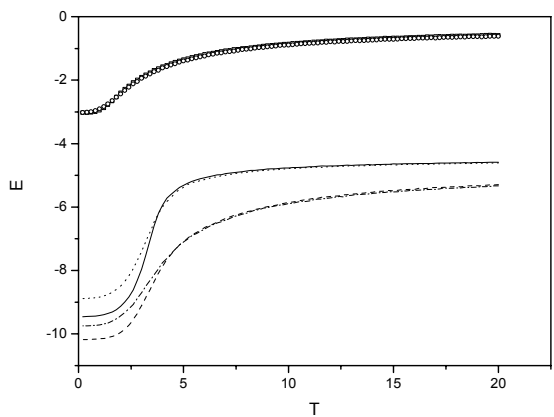

**Figure 1**.- Cooling curves of the mean energy per bead for different systems. Cooling rate: $\Delta T$=-0.1 per 5x10$^3$ MC steps. Solid line: linear chain, $B$=0; dash line: linear chains, $B$=1; dot line: branched chains, $B$=0; dash-dot line: branched chains, $B$=1. Systems without long-range potential, $B$=1, are denoted by symbols. Squares: linear chains, circles: branched chains.

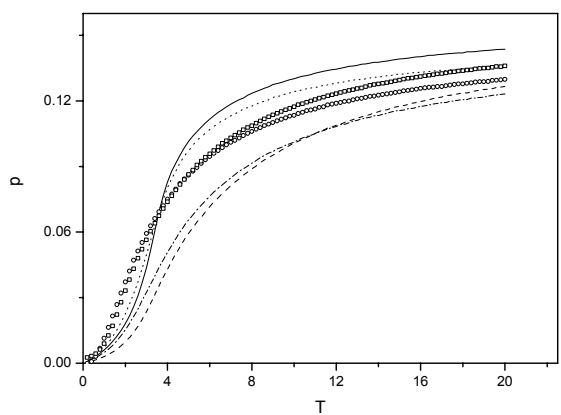

**Figure 2**.- Cooling curves of the mean acceptance probability of a bead jump for different systems. Cooling rate: $\Delta T$=-0.1 per 5x10$^3$ MC steps. Solid line: linear chain, $B$=0; dash line: linear chains, $B$=1; dot line: branched chains, $B$=0; dash-dot line: branched chains, $B$=1. Systems without long-range potential, $B$=1, are denoted by symbols. Squares: linear chains, circles: branched chains.

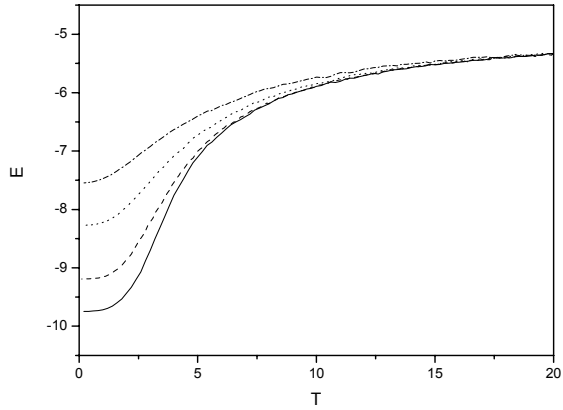

**Figure 3**.- Cooling curves of the mean energy per bead corresponding to branched chains with *B*=1 at different cooling rates. Solid line: $\Delta T$=-0.1 per $5\times10^3$ MC steps; dash line: $\Delta T$=-0.1 per 500 MC steps: dot line: $\Delta T$=-0.1 per 50 MC steps; dash-dot line: $\Delta T$=-0.1 per 5 MC steps.

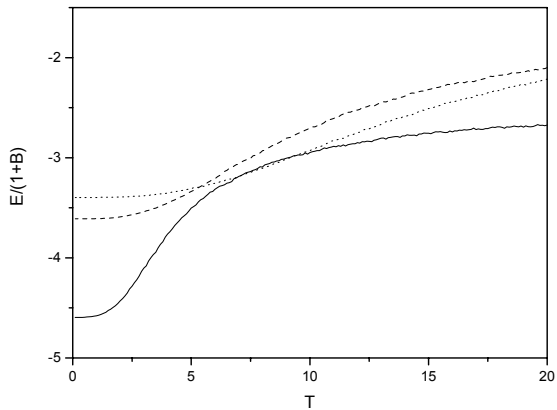

**Figure 4**.- Cooling curves of the reduced mean energy per bead corresponding to branched chains with different values of *B*. Solid line: *B*=1, dash line: *B*=3; dot line: *B*=5. Cooling rate: $\Delta T$=-0.1 per $10^3$ MC steps.

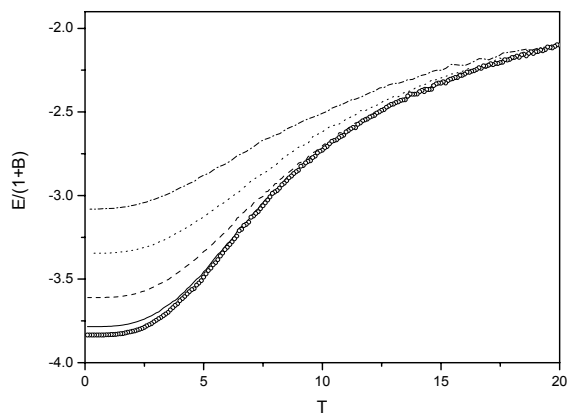

**Figure 5**.- Cooling curves of the reduced mean energy per bead corresponding to branched chains with *B*=3 at different cooling rates. Circles: $\Delta T$=-0.1 per $2.5 \times 10^4$ MC steps; solid line: $\Delta T$=-0.1 per $10^4$ MC steps; dash line: $\Delta T$=-0.1 per $10^3$ MC steps: dot line: $\Delta T$=-0.1 per 50 MC steps; dash-dot line: $\Delta T$=-0.1 per 5 MC steps.

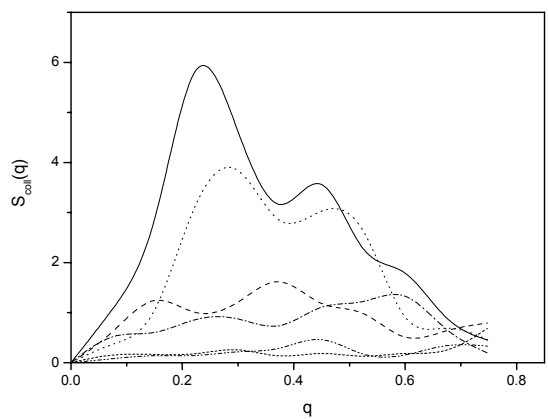

**Figure 6**.- Collective scattering function, $S_{coll}(q)$, for different systems at the lowest temperatures of the cooling process. Solid line: linear chain, *B*=0; dash line: linear chains, *B*=1; dot line: branched chains, *B*=0; dash-dot line: branched chains, *B*=1; short dash line: linear chains, *B*=3; dash-dot-dot line: branched chains, *B*=3.

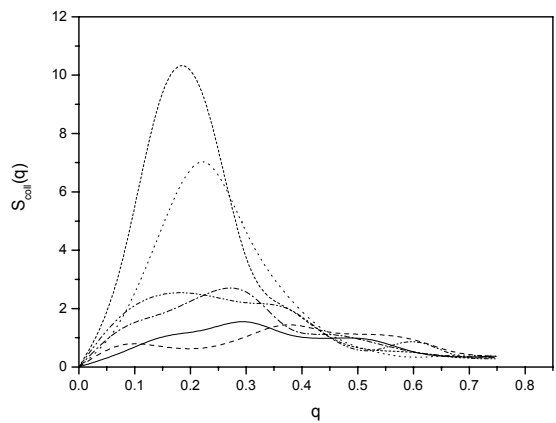

**Figure 7**.- Collective scattering function, $S_{coll}(q)$, at T=3.7, for linear and branched chains with $B$=1, after annealing with several different MC steps. Solid line: linear chains, $10^5$ MC steps; dash line: branched chains, $10^5$ MC steps; dot line: linear chains, $5 \times 10^5$ MC steps; dash-dot line: branched chains, $5 \times 10^5$ MC steps; short dash line: linear chains, $10^6$ MC steps; dash-dot-dot line: branched chains, $10^6$ MC steps.

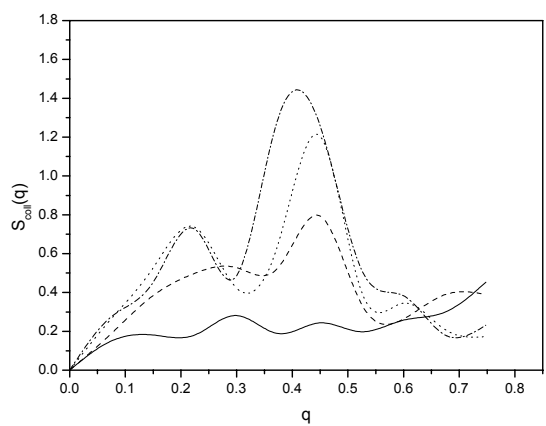

**Figure 8**.- Collective scattering function, $S_{coll}(q)$, at T=3.7, for linear chains with $B$=3, after annealing with several different MC steps. Solid line: $10^5$ MC steps; dash line: $10^6$ MC steps; dot line: $2 \times 10^6$ MC steps; dash-dot line: $3 \times 10^6$ MC steps.

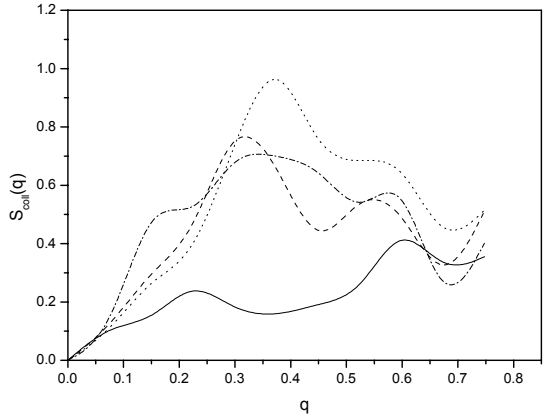

**Figure 9**.- Collective scattering function, $S_{coll}$ ($q$), at T=3.7, for linear chains with alternate bonds of $B$=1 and $B$=5, after annealing with several different MC steps. Solid line: $10^5$ MC steps; dash line: $10^6$ MC steps; dot line: $2\times10^6$ MC steps; dash-dot line: $3\times10^6$ MC steps.

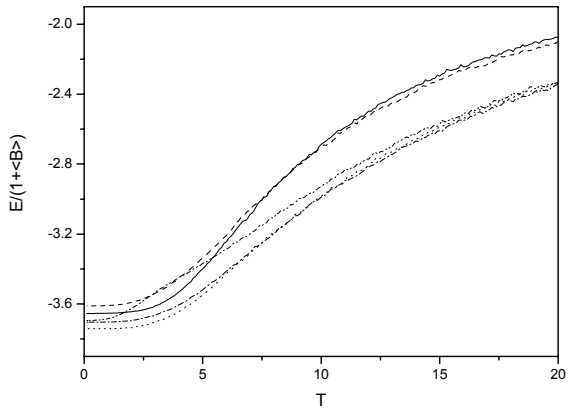

**Figure 10**.- Cooling curves of the reduced mean energy per bead corresponding to different systems with $\langle B\rangle\cong 3$. Solid line: linear chains, $B$=3; dash line: branched chains, $B$=3; dot line: linear chains with alternate bonds of different $B$, $B$=1 and $B$=5; dash-dot: branched chains with $B$=1 for the main chain bonds and $B$=5 for the branch bonds; dash-dot-dot line: branched chains with $B$=5 for the main chain bonds and $B$=1 for the branch bonds. Cooling rate: $\Delta T$=-0.1 per $10^3$ MC steps.

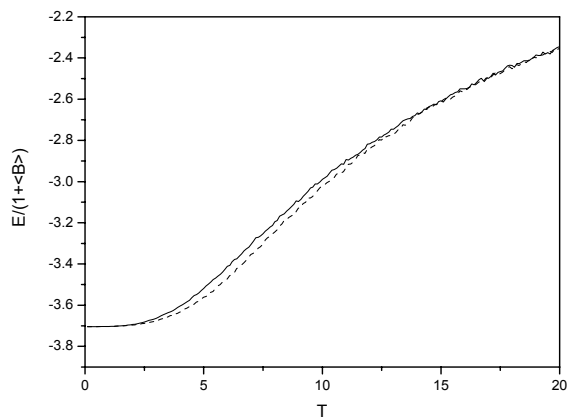

**Figure 11**.- Cooling (solid line) and heating (dash line) curves of the reduced mean energy per bead corresponding to branched chains with $B$=1 for the main chain bonds and $B$=5 for the branch bonds. Rates: $|\Delta T|$=-0.1 per $10^3$ MC steps.

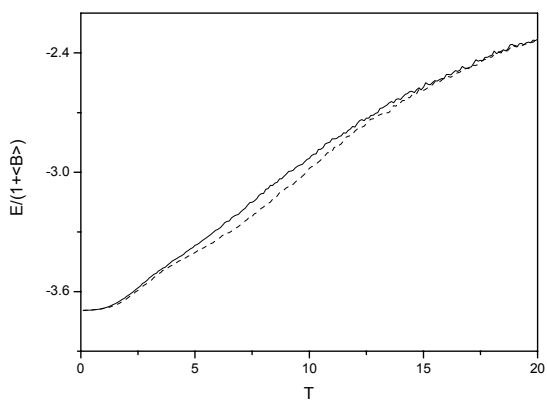

**Figure 12**.- Cooling (solid line) and heating (dash line) curves of the reduced mean energy per bead corresponding to branched chains with $B$=5 for the main chain bonds and $B$=1 for the branch bonds. Rates: $|\Delta T|$=-0.1 per $10^3$ MC steps.